\documentclass[aps,prb,twocolumn,superscriptaddress,showpacs,amsmath,amssymb]{revtex4-1}

\usepackage{amsfonts}
\usepackage{amssymb}
\usepackage{graphicx}
\usepackage{color}

\usepackage{bm}
\usepackage{braket}

\usepackage{hyperref}

\usepackage[normalem]{ulem}

\begin{document}

\title{Fractional Corner Charge of Sodium Chloride}

\author{Haruki Watanabe} \email[]{hwatanabe@g.ecc.u-tokyo.ac.jp}
\affiliation{Department of
Applied Physics, University of Tokyo, Tokyo 113-8656, Japan.}

\author{Hoi Chun Po} \email[]{hcpo@ust.hk}
 \affiliation{Department of Physics, Massachusetts Institute of Technology, Cambridge, Massachusetts 02139, USA.}
\affiliation{Department of Physics, Hong Kong University of Science and Technology, Clear Water Bay, Hong Kong, China}

\begin{abstract}
Recent developments in higher-order topological phases have elucidated on the relationship between nontrivial multipole moments in the bulk and the emergence of fractionally quantized charges at the boundary. Here, we put on the table a proposal of the three-dimensional octupole insulator with fractionally quantized corner charges $\pm e/8$: sodium chloride, commonly known as table salt. 
The fractional quantization of the corner charge is unaffected by the quantum fluctuation of the electric charge per ion, as we demonstrate explicitly with {\it ab initio} calculations. We further show that the electrostatic signature of the fractional charge is well-preserved even when the ideal crystal is sprinkled with defects,
and provide a systematic analysis on how the corner charge contribution could be isolated from the electric-field corrections originating from realistic surface relaxation. Our results suggest that the observation of corner charges from quantized multipole moments in solids might be around the corner.
\end{abstract}

\maketitle

\clearpage

\section{Introduction}
Recent studies of topological phases revealed that insulators can support robust gapless excitations on their boundary due to nontrivial bulk topology. While such gapless excitations reside on the surface in conventional topological insulators,~\cite{RevModPhys.82.3045,RevModPhys.83.1057}
the recently introduced class of higher-order topological insulators feature gapless modes localized to the hinges or corners of the sample.~\cite{schindler2018,song2017,PhysRevB.98.081110,langbehn2017, PhysRevB.98.205129, trifunovic2019,Fang2017,ezawa2018,PhysRevB.98.201114,PhysRevX.9.031003,PhysRevLett.124.036803,Hirayama_honeycomb}

Even when an insulator is completely gapped including its boundary, however, it may still exhibit nontrivial charge signature on its boundary.  
The classic example is the surface charge described by the bulk electric polarization, and the modern theory of electric polarization relates such surface signature with the Berry phase of the bulk band structure.~\cite{PhysRevLett.62.2747,king-smith1993,vanderbilt1993,resta1994,resta2007,lau2016, vanderbilt2018}
When the polarization vanishes, the surfaces will be charge neutral, but then hinges and corners can still be charged.~\cite{benalcazarScience,benalcazar2017,Luka2020,2007.14932,PhysRevResearch.2.033192,PhysRevB.103.035147,He:2020aa,benalcazar2018,PhysRevResearch.1.033074,2007.14932,Kooi2020} The hinge charge and the corner charge are associated to the electric multipole moment of the insulating bulk.~\cite{benalcazarScience,benalcazar2017,kang2018,wheeler2018,ono2019,PhysRevResearch.1.033074,Luka2020,2007.14932,PhysRevResearch.2.033192,PhysRevB.103.035147,He:2020aa,benalcazar2018,PhysRevResearch.1.033074,2007.14932,Kooi2020} In the presence of $n$-fold rotation symmetry, the corner charge for two dimensional systems is quantized to an integer multiple of $e/n$.~\cite{benalcazarScience,benalcazar2017,Luka2020,2007.14932,PhysRevResearch.2.033192,PhysRevB.103.035147,He:2020aa,benalcazar2018,PhysRevResearch.1.033074,2007.14932,Kooi2020} 

To date, experimental realizations of concepts related to the multipole insulators, like the notion of filling anomaly,~\cite{benalcazar2018} have been limited to metamaterials like photonic and phononic systems.~\cite{Serra-Garcia:2018aa, peterson2018, imhof2018,Mittal2019, Haoran2020, PhysRevLett.124.206601,  Peterson1114,Chen2020}
Since such metamaterials platform correspond to weakly interacting bosonic degrees of freedom, they do not realize the multipole insulator as a ground state despite showcasing signatures associated with quantized multipole moments under finite-energy probes. 
In contrast, if a multipole insulator is realized in a solid-state platform using electronic degrees of freedom, it will possess anomalous equilibrium surface charge properties, like fractional hinge or corner charges, which could lead to further technological applications in, for instance, electrochemistry.

In this work, we establish the well-known ionic compound sodium chloride as an octupole insulator with a fractionally quantized corner charge $\pm e/8$.
We also propose a concrete experimental setting for the direct measurement of the corner charge using a non-contact atomic force microscope (AFM).
It should be emphasized that such well-known ionic crystals exemplify the notion of topologically trivial atomic insulators, and, befitting the asserted triviality, the electronic excitation energy spectrum is completely featureless both in the bulk and at the boundaries. Our observations are motivated by a recent work,~\cite{2007.14932} which systematically related the emergence of fractionally quantized boundary charges to the incomplete cancellation of localized integer charges. Since localized integer charges can be simply interpreted as cations and anions, one sees that multipole insulators can have atomic ground states, and are in fact naturally realized in ionic compounds.

\begin{figure}
\begin{center}
\includegraphics[width=\columnwidth]{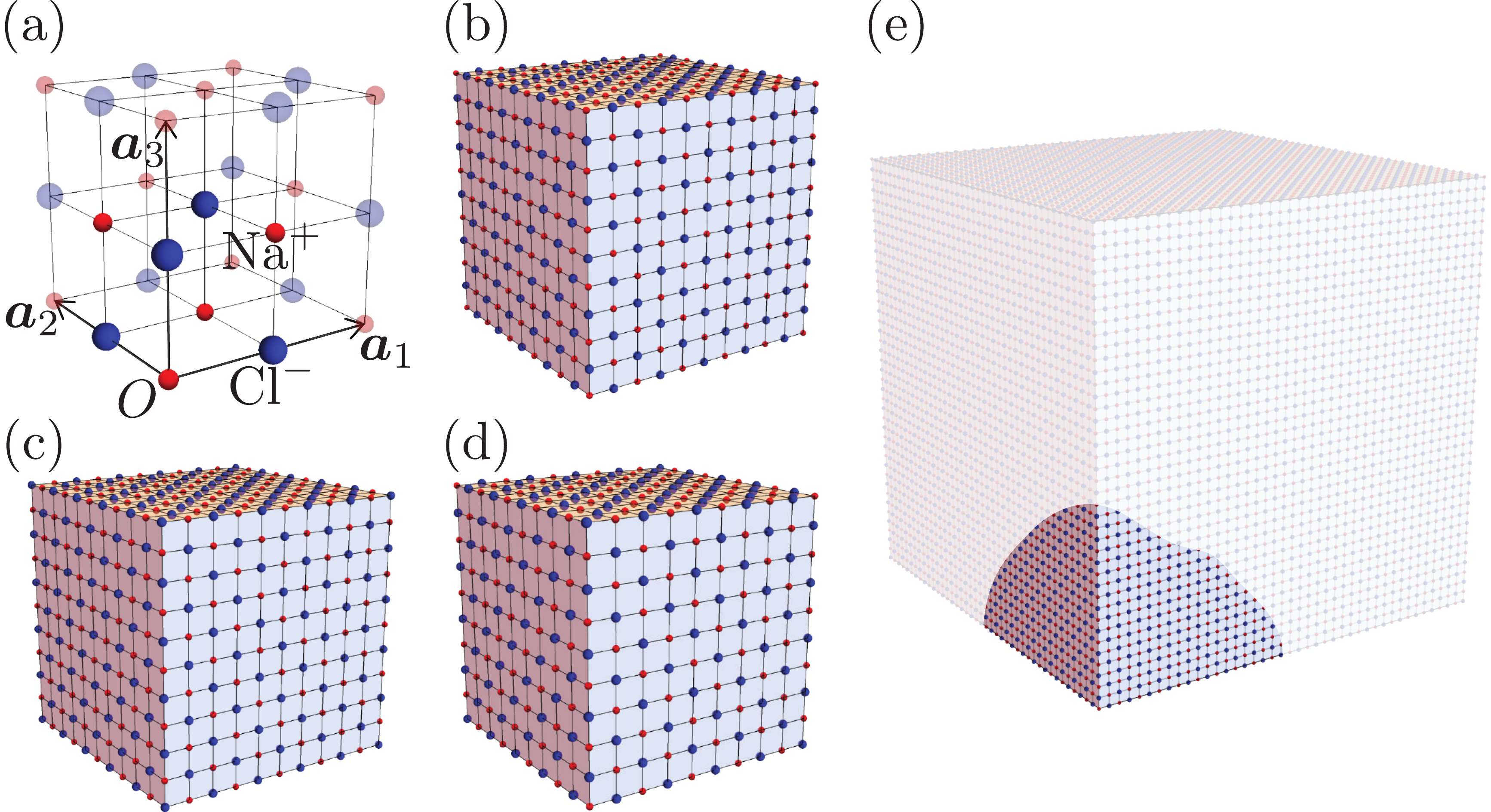}
\caption{\label{figunitcell} 
(a): The conventional unit cell of $\mathrm{Na}\mathrm{Cl}$. $\mathrm{Na}^+$ ions and $\mathrm{Cl}^-$ ions are, respectively, located at Wyckoff position $a$ and $b$ of the space group $Fm\bar{3}m$.  Dimmed ions belong to neighboring cells.
(b,c): An $O$-symmetric cubic sample of $\mathrm{Na}\mathrm{Cl}$ with the side length $L=5a$. All corners are occupied by a $\mathrm{Na}^+$ ion in (b), and by an $\mathrm{Cl}^+$ ion in (c). The total electric charge in the system is $Q=+e$ in (b) and $-e$ in (c).
(d): An $O$-breaking, charge neutral sample of $\mathrm{Na}\mathrm{Cl}$ with the side length $L=4.5a$. 
Corners are occupied by $\mathrm{Na}^+$ ions and $\mathrm{Cl}^+$ ions alternatively.
(e): Illustration of the irrelevance of the entire shape for a charge localized to a corner.
}
\end{center}
\end{figure}

\section{Fractional corner charge of $\mathrm{Na}\mathrm{Cl}$} 
In this section, we establish the emergence of fractional corner charge $\pm e/8$ of $\mathrm{Na}\mathrm{Cl}$ using three independent approaches: the filling anomaly argument, the coupled-wire/layer construction, and an {\it ab initio} tight-binding calculation.

\subsection{Setting}
The space group of the sodium chloride crystal structure is $Fm\bar{3}m$ (225), which belongs to the face-centered cubic system. However,
for the discussions below only the $F432$ (209) subgroup, which can be obtained from $Fm\bar{3}m$ by omitting the inversion symmetry, is essential. Also, as discussed in Ref.~\onlinecite{2007.14932}, in relating the boundary charge signatures with bulk charge distribution one should adopt unit cell conventions which are compatible with the corner. Given the natural cleavage plane of sodium chloride is normal to the $[100]$  (and equivalent) direction, we consider a  conventional unit cell defined by the lattice vectors $\bm{a}_1=(a,0,0)$, $\bm{a}_2=(0,a,0)$, and $\bm{a}_3=(0,0,a)$ to describe a cubic-shaped sample. Note that, in principle, one could consider terminations which are adapted to (some of) the primitive lattice vectors, with which one could also view sodium chloride as a multipole insulator with fractional surface or hinge (i.e., linear) charges. Such terminations, however, are energetically unfavorable and we would instead focus on the natural cubic corner. 

 As illustrated in Fig.~\ref{figunitcell}(a), in each unit cell, four $\mathrm{Na}^+$ ions are located at Wyckoff position $a$: $\bm{r}=(0,0,0)$, $a(0,1,1)/2$, $a(1,0,1)/2$, and $a(1,1,0)/2$ and four $\mathrm{Cl}^-$ ions are at the Wyckoff position $b$: $\bm{r}=a(1,1,1)/2$, $a(1/2,0,0)$, $a(0,1/2,0)$, and $a(0,0,1/2)$. The point group symmetry of the crystal structure we would need is the orientation-preserving octahedral symmetry $O$, which is composed of the four fold rotation symmetry about each axis and the three fold rotation about the $(1,1,1)$ axis. Note that while the crystal also has the inversion symmetry, we would not need it in the argument. The site symmetry of the Wyckoff position $a$ and $b$ are both $O$.

\subsection{Filling anomaly} 

As shown recently,~\cite{2007.14932} the corner charge $Q_c$ of a cubic sample can be predicted from the bulk charge distribution as
\begin{align}
Q_c=\frac{1}{8}q_a=\frac{1}{8}q_b\mod \frac{1}{4}e.
\label{formula0}
\end{align}
Here, $q_w$ ($w=a,b$) represents the sum of electric charges of Wannier orbitals centered at the Wyckoff position $w$ and ionic charges localized to the Wyckoff position in the bulk. By definition, $q_w$ is quantized to an integer. This formula assumes that the electric charge density, the electric polarization, and the electric quadrupole moment, defined with respect to the conventional lattice vectors, all vanish in the bulk of the system.  We also assume that the system does not have nontrivial topological invariants and therefore can be described as an atomic insulator. These assumptions are satisfied in the ionic crystals with the sodium chloride structure as we discuss below. Furthermore, the surface termination of the cubic sample must be chosen in such a way that the octahedral symmetry about the center of the cube is preserved, as illustrated in Fig.~\ref{figunitcell}(b,c).   Since the Wyckoff position $a$ and $b$ are, respectively, occupied by a $\mathrm{Na}^+$ ion and a $\mathrm{Cl}^{-}$ ion in the current example, we have $q_a=+e$ and $q_b=-e$. Therefore,
\begin{align}
Q_c=\frac{1}{8}e\mod \frac{1}{4}e.
\label{formula1}
\end{align}

The $e/4$-ambiguity in Eqs.~\eqref{formula0} and \eqref{formula1} suggests that there is no way to predict whether the corner charge is $+e/8$ or $-e/8$ from the bulk perspective even in the presence of the octahedral symmetry. Indeed, the corner charge is $+e/8$ for the surface termination in Fig.~\ref{figunitcell}(b) in which all eight corners are occupied by an $\mathrm{Na}^+$ ion. When the cutting position is slightly shifted and all corners are occupied by a $\mathrm{Cl}^-$ ion instead, the corner charge becomes $-e/8$ without affecting the bulk of the system at all.

The fractional corner charge $\pm e/8$ can be most easily understood based on the idea of ``filling anomaly"~\cite{benalcazar2018}, which concerns the total electric charge $Q_{\text{tot}}$ in the system.  When the bulk, surface, and hinges are all charge neutral, $Q_{\text{tot}}$ must be distributed equally to the eight corners of the cube as a consequence of the octahedral symmetry. Therefore, each corner must carry the charge
\begin{align}
Q_c=\frac{1}{8}Q_{\text{tot}}.
\label{formula2}
\end{align}
This formula does not contain any ambiguity  because it assumes the detailed information about the surface termination not only about the charge distribution in the bulk.

In our current example, when the side length of the cube is $L=Na$ and the surface termination is as illustrated in Fig.~\ref{figunitcell}(b), the total number of $\mathrm{Na}^+$ ions in the system is $N_+=(N+1)^3+3(N+1)N^2= 4 N^3+6N^2+3N+1$ and that of $\mathrm{Cl}^-$ ions is $N_-=N^3+3N(N+1)^2=4 N^3+6N^2+3N$. Hence the total electric charge of the cube is $Q_{\text{tot}}=e(N_+-N_-)=+e$, which implies $Q_c=+e/8$.  For the surface termination in Fig.~\ref{figunitcell}(c), $\mathrm{Na}^+$ ions and $\mathrm{Cl}^-$ ions are interchanged and thus $Q_c=-e/8$.  More generally, the octahedral symmetry relates the total electric charge in the system $Q_{\text{tot}}$ to the local electric charge $q_w$ as
\begin{align}
Q_{\text{tot}}=q_{a}=q_{b}\mod 2e.
\label{formula3}
\end{align}
The second equality assumes that the polarization and the quadrupole moment vanish in the bulk.
Plugging Eq.~\eqref{formula3} into Eq.~\eqref{formula2}, one reproduces Eq.~\eqref{formula0}. 

\begin{figure}[t]
\begin{center}
\includegraphics[width=\columnwidth]{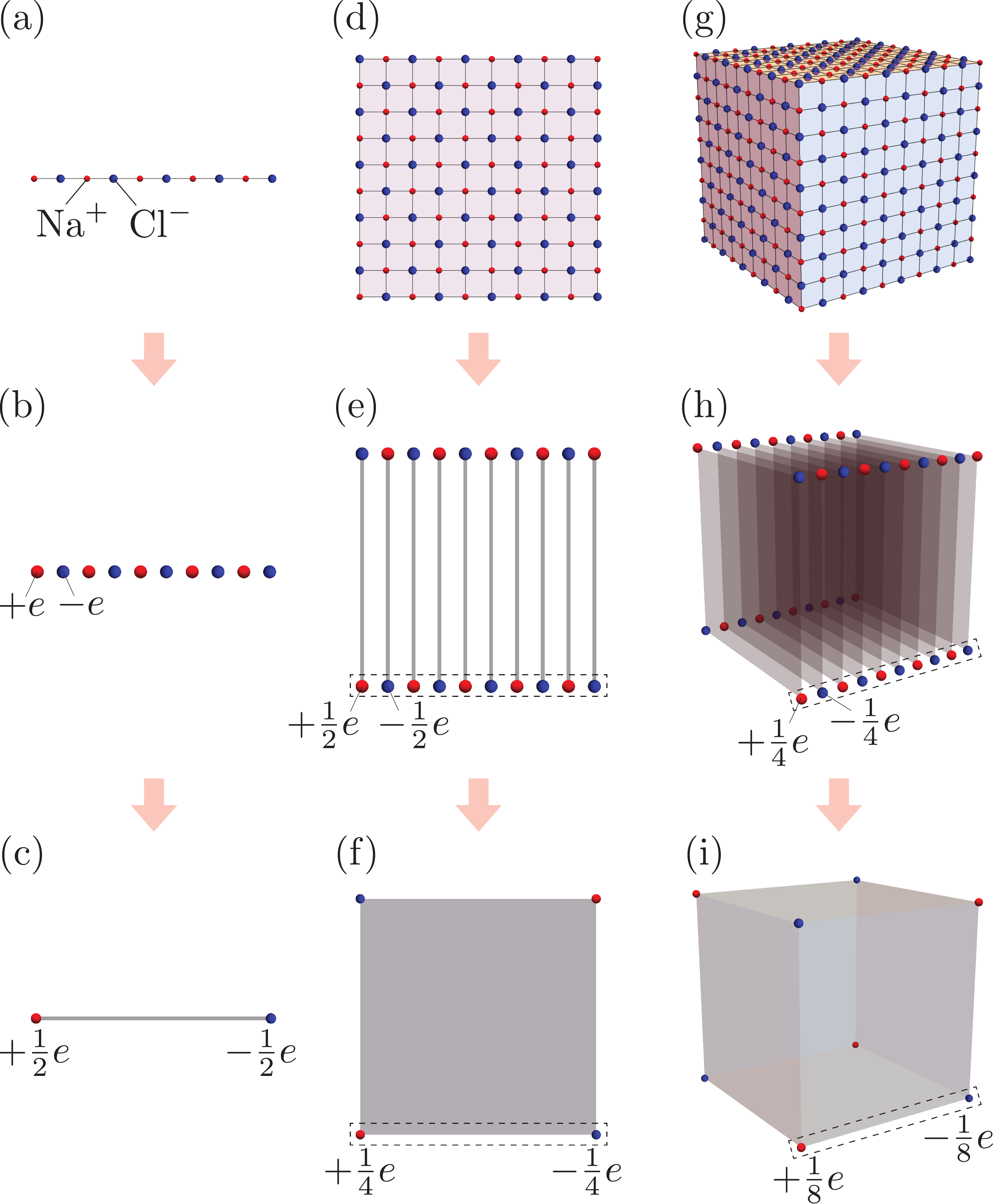}
\caption{\label{figwire} 
Understanding of the fractional corner charges from the coupled-wire/layer construction.
(a)--(c): The wire of $\mathrm{Na}\mathrm{Cl}$ exhibits $\pm e/2$ charges at ends as a result of the polarization $+e/2$. 
(d)--(f): The monolayer of $\mathrm{Na}\mathrm{Cl}$ may be regarded as a periodic array of $\mathrm{Na}\mathrm{Cl}$ wires. The end charges $\pm e/2$ in the gray dashed box can be regarded as the 1D system illustrated in (a). They produce the $\pm e/4$ corner charge of the monolayer.
(g)--(i): Similarly, the cubic crystal of $\mathrm{Na}\mathrm{Cl}$ may be regarded as a periodic array of $\mathrm{Na}\mathrm{Cl}$ layers. The corner charges $\pm e/4$ in the gray dashed box can be regarded as the 1D system illustrated in (a). They produce the $\pm e/8$ corner charge of the cubic sample.
}
\end{center}
\end{figure}

\subsection{Coupled wire/layer construction} 
\label{sec:coupled}
The filling anomaly argument presented above relies on the perfect octahedral symmetry of the sample and is effective only when the total electric charge $Q_{\text{tot}}$ is nonzero. 
In reality, however, the octahedral symmetry may be broken and the electric charge might be neutralized (i.e., $Q_{\text{tot}}=0$), as illustrated in Fig.~\ref{figunitcell}(d).
To understand the corner charge of $\mathrm{Na}\mathrm{Cl}$ even in that case, let us present a coupled-wire/layer type argument. 
To this end, we theoretically imagine the one-dimensional and two-dimensional version of $\mathrm{Na}\mathrm{Cl}$, as illustrated in Fig.~\ref{figwire}(a,d).
In one dimension, a $\mathrm{Na}\mathrm{Cl}$ wire in Fig.~\ref{figwire}(a) can be understood as an alternating arrangement of point charges $\pm e$ as in Fig.~\ref{figwire}(b).  Reflecting the bulk polarization $+ea/2$ of the 1D system, which is quantized by any of the inversion, mirror, or two-fold rotation symmetries flipping the long direction of the chain,~\cite{PhysRevLett.62.2747,king-smith1993,vanderbilt1993,resta1994,resta2007,lau2016, vanderbilt2018} the two ends of the wire exhibit the fractional charge $\pm e/2$ as depicted in Fig.~\ref{figwire}(c). 
We remark that the emergence of fractional edge charges in this hypothetical one-dimensional crystal of $\mathrm{Na}\mathrm{Cl}$ is closely related to the conducting domain wall found in polyacetylene~\cite{jackiw1976, su1979, heeger1988}. More details on the connection can be found in Appendix \ref{app:1DNaCl}, which also contains, as a precursor to our discussion in Sec.\ \ref{sec:E_corner},  the analytical form of the electric field near the edge of a semi-infinite chain of alternating point charges.

Next, in two dimensions, a $\mathrm{Na}\mathrm{Cl}$ monolayer in Fig.~\ref{figwire}(d) can be decomposed into a periodic array of $\mathrm{Na}\mathrm{Cl}$ wires with alternating polarization as in Fig.~\ref{figwire}(e). The bulk of the $\mathrm{Na}\mathrm{Cl}$ layer does not have the net polarization since the polarization of two neighboring wires cancel with each other. The $\pm e/2$ charges at the end of wires, shown in the gray dashed box in Fig.~\ref{figwire}(e), can be regarded as the 1D system in Fig.~\ref{figwire}(b), but the charge unit is effectively halved into $e/2$. Hence, the charges appearing at the end of this 1D system are $\pm e/4$, which account for the corner charge $\pm e/4$ of the $\mathrm{Na}\mathrm{Cl}$ monolayer in Fig.~\ref{figwire}(f). In other words, the bulk of the 2D system possesses the quadrupole moment $+e/4$.

Finally, in three dimensions, a $\mathrm{Na}\mathrm{Cl}$ cube in Fig.~\ref{figwire}(g) may be regarded as a periodic array of $\mathrm{Na}\mathrm{Cl}$ layers with alternating quadrupole moment as in Fig.~\ref{figwire}(h). This construction proves the lack of the net polarization and the quadrupole moment in the bulk of the 3D system. Moreover, the effective 1D wire made of the corner charge $\pm e/4$ of each layer [the gray dashed box in (h)] explains the corner charge $\pm e/8$ of the $\mathrm{Na}\mathrm{Cl}$ cube illustrated in Fig.~\ref{figwire}(i).    In this surface termination of the cube, four of the eight corners are occupied by a $\mathrm{Na}^+$ ion and the other four are occupied by a $\mathrm{Cl}^-$ ion, indicating that the octahedral symmetry is explicitly broken. 
The sign of the corner charge alternates accordingly and the octupole moment of $\mathrm{Na}\mathrm{Cl}$ is manifest. Even though the octahedral symmetry is broken and Eq.~\eqref{formula2} is not applicable to the sample in Fig.~\ref{figwire}(g), the corner charge is still quantized to $\pm e/8$ in the coupled-wire/layer construction. 

This result is well-anticipated because the corner charge is a local property of each corner and the entire shape of the crystal is irrelevant.
More generally, as far as the crystal structure around a corner is the same as the $O$-symmetric case, even for only a fraction of a cube illustrated in Fig.~\ref{figunitcell}(e), the amount of the charge localized to a corner should be the same as in the $O$-symmetric crystal.

\subsection{Quantum charge fluctuation \label{sec:TB}} 
In the above coupled-wire/layer construction, we treated each ion as a point charge with quantized electric charge $\pm e$.
In reality, however, the actual electric charge assigned to each ion is not quantized due to quantum fluctuations. Here we demonstrate that the corner charge remains quantized to $\pm e/8$ even in the presence of such quantum fluctuations by numerically investigating an {\it ab initio} four-band tight-binding model. 

\begin{figure}[t]
\begin{center}
\includegraphics[width=\columnwidth]{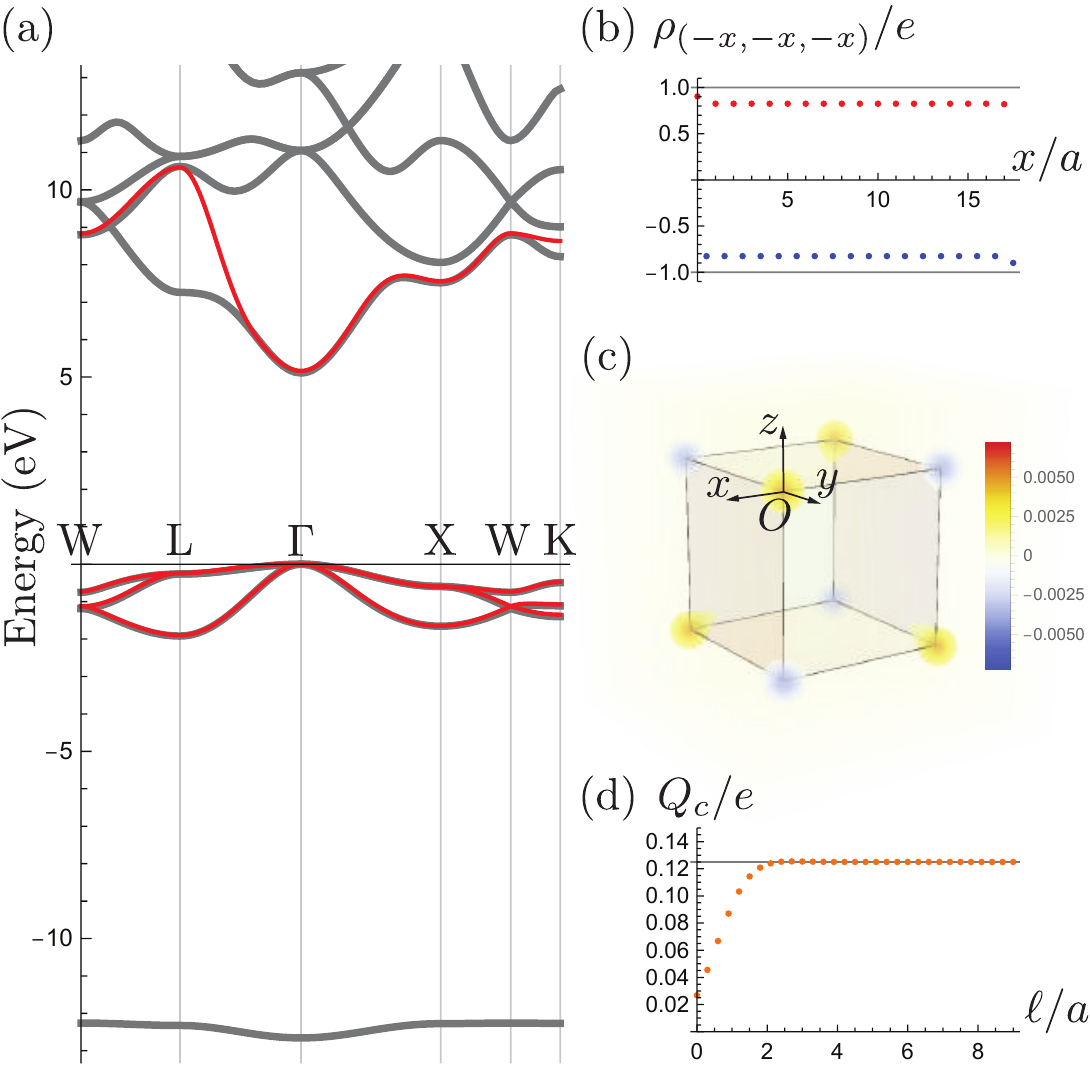}
\caption{\label{figTB} 
(a): The band structure of our tight binding model (red) and the first-principles calculation (gray).
(b): The charge distribution $\rho_{\bm{r}}$ in the diagonal direction.
(c): The coarse-grained charge density $\tilde{\rho}_{\text{tot}}(\bm{r})$.
(d): The corner charge $Q_c$ in Eq.~\eqref{defQc} as a function of $\ell$.
}
\end{center}
\end{figure}

Our model is composed of $N_{\text{orb}}=4$ orbitals per primitive unit cell: the 3$s$ orbital of Na atoms and the three 3$p$ orbitals $p_x,p_y,p_z$ of Cl atoms. When all of these orbitals are unoccupied, an Na ion and a Cl ion respectively have the charge $+e$ and $+5e$. The hopping parameters of the tight-binding model are obtained from band calculations based on the density functional theory assuming the spin-rotation symmetry. See Fig.~\ref{figTB}(a) for the bulk band structure described by our model.

For the purpose of computing the corner charge, we assume a cubic shape with the side length $L=17.5a$ with the open boundary condition. The total number of Na ions and Cl ions in the system is, respectively, $N_{\text{uc}}\equiv 4\times 18^3$, and the filling anomaly argument is not directly applicable.  By diagonalizing the single-particle Hamiltonian, we find in total $2\times N_{\text{orb}}\times N_{\text{uc}}$ energy levels including the spin degeneracy. The charge-neutral ground state $|\Phi\rangle$ is given by the Slater determinant of the lowest $2\times3\times N_{\text{uc}}$ states. 

The electric charge of a Na site $\bm{r}\in\Lambda_a$ is given by
\begin{equation}
\rho_{\bm{r}}=e-e\sum_{\sigma=\uparrow,\downarrow}\langle \Phi|\hat{a}_{\bm{r},\sigma}^\dagger\hat{a}_{\bm{r},\sigma}|\Phi\rangle
\end{equation}
and that of a Cl site $\bm{r}\in\Lambda_b$ is given by 
\begin{equation}
\rho_{\bm{r}}=5e-e\sum_{o=p_x,p_y,p_z}\sum_{\sigma=\uparrow,\downarrow}\langle \Phi|\hat{b}_{\bm{r},o,\sigma}^\dagger\hat{b}_{\bm{r},o,\sigma}|\Phi\rangle,
\end{equation}
where $\hat{a}_{\bm{r},\sigma}$ and $\hat{b}_{\bm{r},o,\sigma}$ are, respectively, the annihilation operator of the Na $s$ orbital and Cl $p$ orbitals.  We plot the charge distribution $\rho_{\bm{r}}$ of the ground state along the diagonal direction in Fig.~\ref{figTB}(b).
As one can see, the system size $L=17.5a$ is sufficiently large to take into account the surface effect.  We find $\rho_{\bm{r}}/e=+0.8248\cdots$ for the Na site and $\rho_{\bm{r}}/e=-0.8248\cdots$ for the Cl site in the bulk of the sample, both of which are significantly reduced from the ideal value $\pm e$ due to the quantum fluctuations.

To extract the corner charge of the cubic sample, we introduce a coarse-grained charge density $\tilde{\rho}(\bm{r})$ by a Gaussian convolution (see Sec.~6.6 of \onlinecite{jackson2007classical}):
\begin{align}
&\tilde{\rho}_{\text{tot}}(\bm{r})\equiv\sum_{\bm{r}'\in\Lambda_a} G(\bm{r}-\bm{r}')\rho_{\bm{r}'}+\sum_{\bm{r}'\in\Lambda_b} G(\bm{r}-\bm{r}')\rho_{\bm{r}'},\label{coarsegrain}\\
&G(\bm{r})\equiv (2\pi\lambda^2)^{-3/2}e^{-\frac{|\bm{r}|^2}{2\lambda^2}}.
\end{align}
Here, $\Lambda_w$ is the set of lattice points belonging to the Wyckoff position $w=a,b$ in the cubic sample. 
The parameter $\lambda$ should be set sufficiently large so that $\tilde{\rho}_{\text{tot}}(\bm{r})$ neglects the microscopic details of the charge distribution.  
In particular, $\tilde{\rho}(\bm{r})$ should be negligibly small in the bulk.  In this work, we set $\lambda=a$  for which we find $|\tilde{\rho}_{\text{tot}}(\bm{r})/e|<10^{-13}$ in the bulk.
We plot $\tilde{\rho}_{\text{tot}}(\bm{r})$ obtained thereby in Fig.~\ref{figTB}(c), which shows a peak around each corner of the cube.  We determine the corner charge $Q_c$ by the spatial integral of $\tilde{\rho}_{\text{tot}}(\bm{r})$ over a semi-infinit box that covers a single peak:
\begin{align}
Q_c=\int_{-\ell}^\infty dx\int_{-\ell}^\infty dy\int_{-\ell}^\infty dz\tilde{\rho}_{\text{tot}}(\bm{r}).\label{defQc}
\end{align}
We plot $Q_c$ as a function of $\ell$ in Fig.~\ref{figTB}(d), which converges quickly to $e/8$ when $\ell\gg a$. For example, $Q_c/e=0.124999998$ at $\ell=6a$.

\section{Electric field around a corner \label{sec:E_corner}}
Having established the emergence of the fractional corner charge, we next examine the electric field around the corner of a cubic crystal of NaCl and discuss its possible measurement using AFM.
We consider a cubic sample with the side length $L$ and focus on the corner on the top surface, which we set the origin of the coordinate. We assume the corner is occupied by a $\mathrm{Na^+}$ ion and hence supports the corner charge $Q_c=+e/8$. In this section, we mostly focus on the simple model in which the charge density of each ion is given by the delta function $\pm e\delta^3(\bm{r})$.
An exception is Sec.\ \ref{sec:distort}, in which we investigate analytically the effect of surface distortion on the electric field near a corner, and our analytical results there apply even in the presence of the quantum fluctuation discussed in Sec.\ \ref{sec:TB}. 
Although our numerical results in this section are based on the point-charge model, as we argue below the quantum fluctuation would only impact a small correction to them.

\subsection{Contribution from point charges}
\label{sec:pc}
\begin{figure}[t]
\begin{center}
\includegraphics[width=\columnwidth]{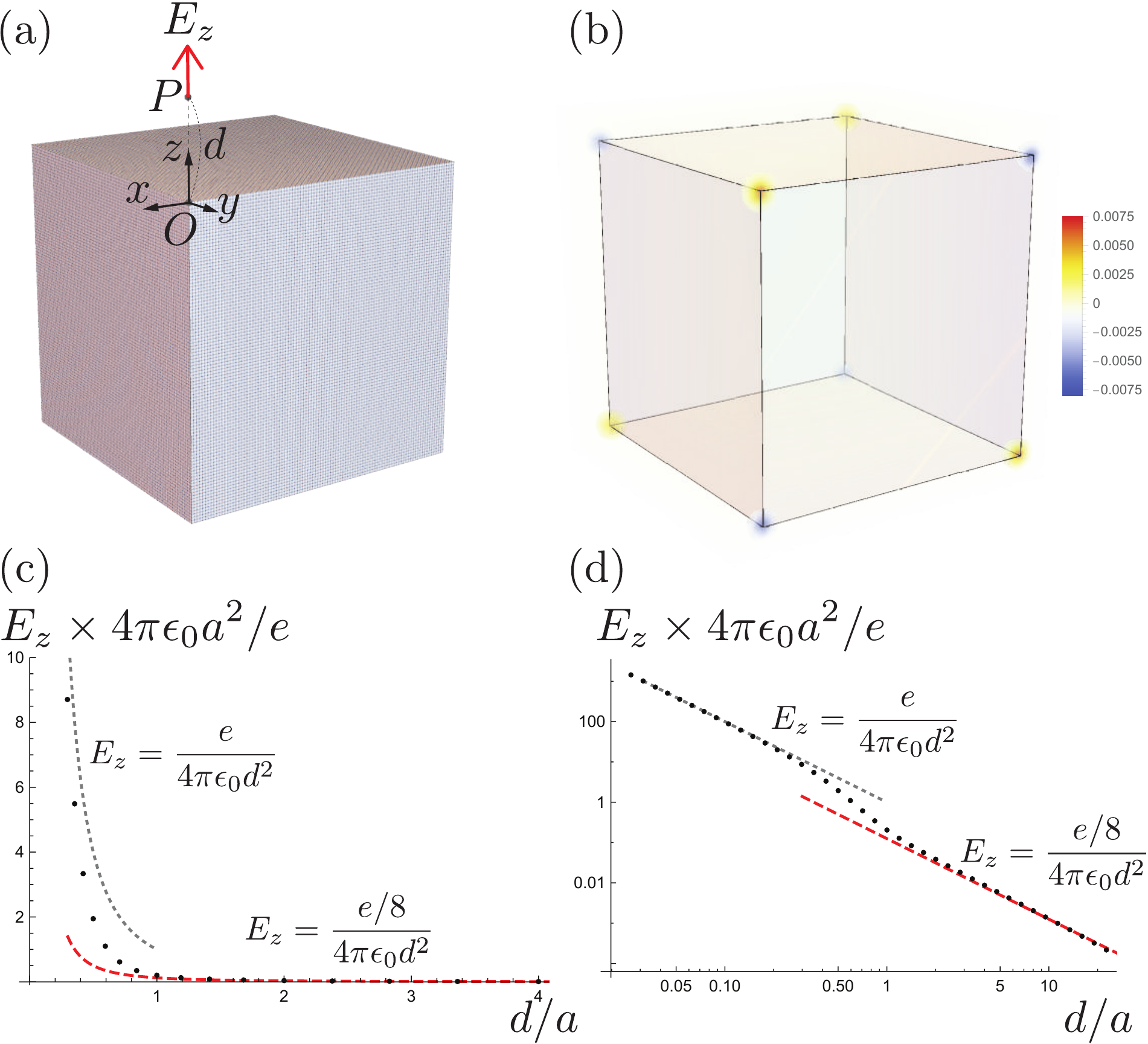}
\caption{\label{figcorner} 
(a): A cubic sample of $\mathrm{Na}\mathrm{Cl}$ with side length $L= 49.5a$. 
(b): The coarse-grained charge density $\tilde{\rho}_{\text{tot}}(\bm{r})$ for the sample in (a). 
(c): The $z$ component of the electric field $E_z$ at distance $d$ above the corner $O$.  As $d$ increases, $E_z$ changes from the contribution of the single $\mathrm{Na}^+$ ion (the gray dotted curve) to the contribution of the corner charge $Q_c=+e/8$ (the red dashed curve).  (d) The log-log plot of (c).
}
\end{center}
\end{figure}

In the point-charge model, the total charge density of the cubic sample is
\begin{align}
\rho_\text{tot}(\bm{r})&=\sum_{\bm{r}_a\in\Lambda_a}e\delta^3(\bm{r}-\bm{r}_a)-\sum_{\bm{r}_b\in\Lambda_b}e\delta^3(\bm{r}-\bm{r}_b).
\label{totalcharge}
\end{align}
We plot the coarse-grained charge density $\tilde{\rho}_{\text{tot}}(\bm{r})\equiv\int d^3r'G(\bm{r}-\bm{r}')\rho_\text{tot}(\bm{r}')$ with $\lambda=a$ for the cube with the side length $L=49.5a$ in Fig.~\ref{figcorner}(b), which is nonzero only around corners of the cube. The bulk, surfaces, and hinges are all charge neutral. 

\begin{figure}[t]
\begin{center}
\includegraphics[width=\columnwidth]{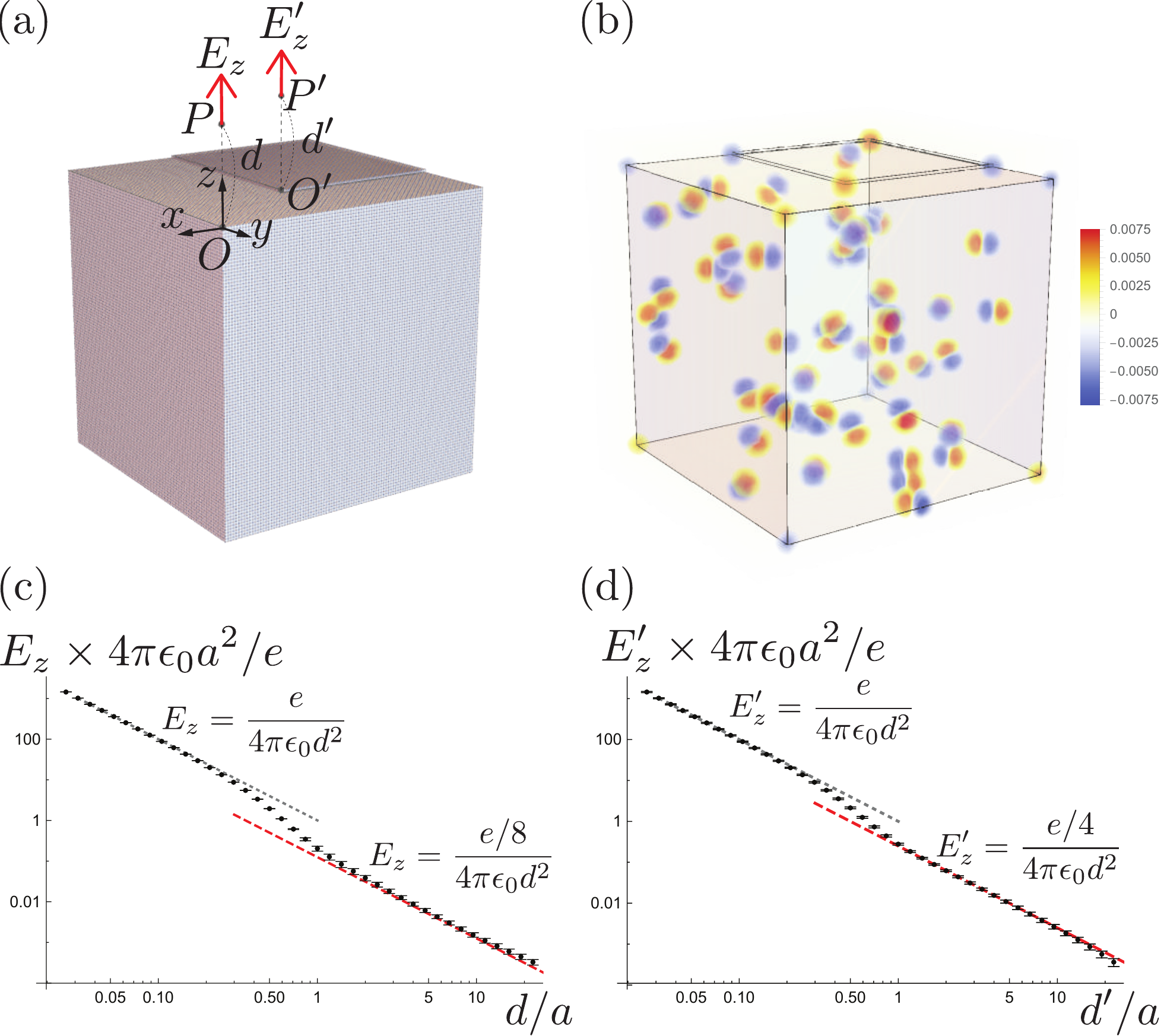}
\caption{\label{figexp} 
(a): A step is introduced on the top surface of the cubic sample in Fig.~\ref{figcorner}(a). The corner $O'$ of the step edge is at $\bm{r}=(-20, -14.5, 0.5)a$. 
(b): The coarse-grained charge density $\tilde{\rho}_{\text{tot}}(\bm{r})$ for the sample in (a).  50 dipolar defects are included in the calculation.  
(c,d): The $z$ component of the electric field $E_z$  and $E_z'$ at $P$ and $P'$, respectively. 
The error bars are obtained as standard deviation over $1000$ disorder realizations. The small deviation of the data points from the red line around $d\gtrsim 10a$
can be explained by the corner charges from both the corners of the cube and those of the step.}
\end{center}
\end{figure}

The electric field produced by these point charges is given by
\begin{equation}
\bm{E}(\bm{r})=\frac{e}{4\pi\epsilon_0}\sum_{\bm{r}_a\in\Lambda_a}\frac{\bm{r}-\bm{r}_a}{|\bm{r}-\bm{r}_a|^3}-\frac{e}{4\pi\epsilon_0}\sum_{\bm{r}_b\in\Lambda_b}\frac{\bm{r}-\bm{r}_b}{|\bm{r}-\bm{r}_b|^3}.
\end{equation}
We are interested in the behavior of $\bm{E}(\bm{r})$ as a function of the distance $d\equiv|\bm{r}|$ from the corner [see Fig.~\ref{figcorner}(a)]. By computing the sum numerically, we find that the electric field behaves as
\begin{equation}
\bm{E}(\bm{r})\sim\frac{e/8}{4\pi \epsilon_0}\frac{\bm{r}}{d^3}.
\label{Fz1}
\end{equation}
at an intermediate length scale $a<d\ll L$, which can be interpreted as the contribution from the corner charge $Q_c=e/8$.
On the other hand, at a shorter distance $d \lesssim a$, the electric field is dominated by the $\mathrm{Na^+}$ ion at the corner:
\begin{equation}
\bm{E}(\bm{r})\sim\frac{e}{4\pi \epsilon_0}\frac{\bm{r}}{d^3}.
\label{Fz2}
\end{equation}
The crossover from the behavior in Eq.~\eqref{Fz1} to the one in Eq.~\eqref{Fz2} occurs around $d\sim a$.
For example, our results for $\bm{r}=(0,0,d)$ is shown in Fig.~\ref{figcorner}(c,d).

\subsection{Robustness against disorders}

We further assess the robustness of the above result by considering the effects of point defects in the crystal. 
The defects could come from the intrinsic thermal excitations of the ions, leading to Schottky defects formed by a pair of cation and anion vacancies, or from extrinsic aliovalent impurities like the substitution $\mathrm{Na}^+ \rightarrow \mathrm{Ca}^{2+}$ together with the associated cation vacancy. Importantly, isolated point defects in an ionic crystal are electrically charged, and the strong Coulomb force binds them into small dipoles with dipole moments on the order of $p \sim ea$ (a process that is also called ``defect association'' in the literature~\cite{PhysRev.126.1367}). The binding energy is typically at the $0.1-1$ eV scale, which implies the dipolar picture for defects is accurate for samples at room temperature or below, provided proper annealing procedures were used~\cite{PhysRev.126.1367}. 
As the dipolar field falls off more rapidly as $\sim p/d^3$ compared with the corner charge contribution of $\sim e/d^2$, the presence of such defects has a negligible effect on the $1/d^2$ term of the electric field. This is demonstrated in Fig.~\ref{figexp}(c,d), in which we include in the calculation 50 dipolar defects (corresponding to a defect concentration of $100$ ppm) of dipole moment $p = e a$ randomly oriented along the primitive lattice vectors. The signature of the $e/8$ corner charge clearly remains observable.

A more interesting form of defects concerns step edges on the surface of the crystal.  In fact, one additional layer of atoms on top of a three-dimensional bulk behaves effectively as a two-dimensional  $\mathrm{Na}\mathrm{Cl}$ monolayer, which, as we have discussed in Sec.~\ref{sec:coupled}, realizes a quadrupole insulator with $\pm e/4$ corner charges [Fig.~\ref{figexp}(d)]. When a corner of the step edges is at a distance $d_{OO'}$ away from the corner of the three-dimensional bulk, it could lead to deviation of $E_z$ from the ideal value a distance  $d \sim d_{OO'}$. This deviation can be seen in Fig.~\ref{figexp}(c). 

\subsection{Contribution from surface distortion \label{sec:distort}}
\begin{figure}[t]
\begin{center}
\includegraphics[width=\columnwidth]{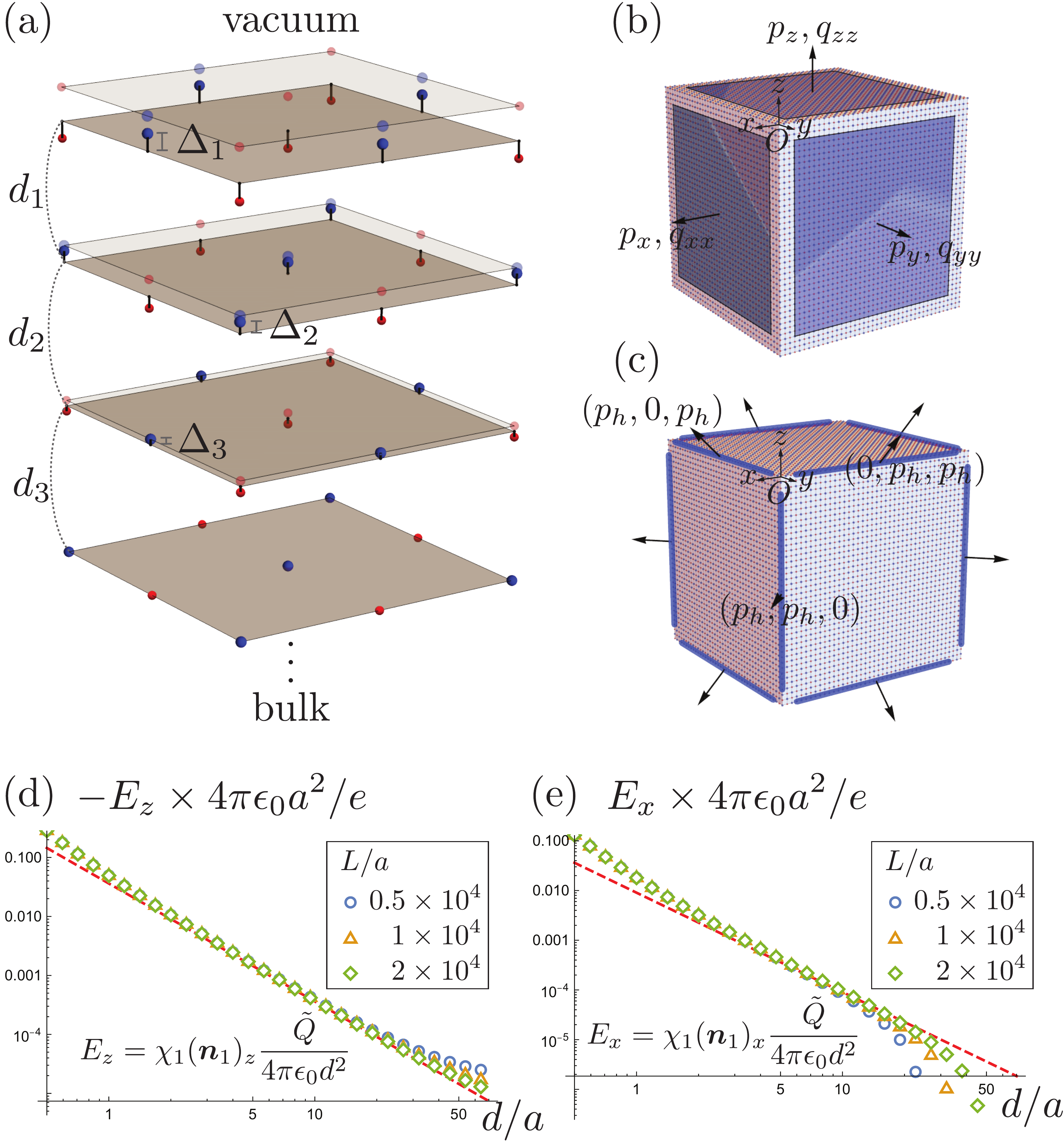}
\caption{\label{figqtilde} 
(a): The surface relaxation of the $(001)$ surface of NaCl.  The scales are exaggerated for illustration.
(b): Out-of-plane dipole moments and quadrupole moments induced near surface.
(c): Dipole moments induced on hinges.
(d): The contribution to the $z$-component of the electric field from the surface relaxation. We use the experimental values~\cite{VOGT2001155} of $\Delta_1=0.07$~\AA, $\Delta_2=0.01$~\AA, $\Delta_3=0$, $d_1=2.76$~\AA, $d_2=2.80$~\AA, and $d_3=2.81$~{\AA}, which give $\tilde{Q}=-0.036e$.  The red dashed line represents the analytic expression in Eqs.~\eqref{surface001}.
(e): The same as (d) but for the $x$-component of the electric field.}
\end{center}
\end{figure}

In any finite-size crystals with boundaries, the lattice structure may be distorted near surfaces. For instance, the top few layers of a crystal may contract inwards, which induce a non-zero surface dipole density when compared against the non-relaxed structure. Similarly, other motion of the surface atoms could also lead to higher multiple moments at the boundary.  For sodium chloride, experiments~\cite{KASHIHARA1989477,ROBERTS199975,VOGT2001155} found that the inter-layer distance is reduced near the surface and that sodium ions and chloride ions, respectively, move inward and outward within a layer, as illustrated in Fig.~\ref{figqtilde}(a).  The above calculation of the corner charge and the electric field around a corner did not take into account such surface effects.  Here we discuss their consequences assuming that the lattice distortion does not spontaneously lower the symmetry of the finite crystal, e.g., the distorted faces are still related by a $C_3$ symmetry. We remark that  as sodium chloride is strongly ionic and we consider its natural cubic cleavage, its surfaces are chemically and structurally stable. The only important surface distortion is the elastic relaxation of ion positions, which we address below. Such lattice distortion near surfaces has no impact on the fractionally quantized corner charge and does affect the $1/d^2$ term of the electric field around corners, as we discuss both analytically and numerically below. 

Let $\bm{r}_0$ be a lattice site of an undistorted cubic crystal and $\rho_{\bm{r}_0}(\bm{r})$ be the charge distribution associated with the ion at $\bm{r}_0$. We have
\begin{align}
\int d^3r\,\rho_{\bm{r}_0}(\bm{r}+\bm{r}_0)=\pm e,
\end{align}
depending on whether $\bm{r}_0$ is the $\mathrm{Na}$ site or $\mathrm{Cl}$ site. 
These point charges of the undistorted lattice generate the $\bm{r}/d^3$-type electric field of the corner charge discussed in Sec.~\ref{sec:pc}.

To examine the effect of distortion, we define the dipole moment and quadrupole moment by
\begin{align}
p_i^{\bm{r}_0}&=\int d^3r\,r_i\rho_{\bm{r}_0}(\bm{r}+\bm{r}_0),\\
q_{ij}^{\bm{r}_0}&=\int d^3r\,(r_ir_j-\delta_{ij}r^2/3)\rho_{\bm{r}_0}(\bm{r}+\bm{r}_0).
\end{align}
Higher multipole moments are irrelevant for the $1/d^2$ term of the electric field around a corner.  
We stress that the multipole expansion here takes into account both the shape of the Wannier function (from the charge density $\rho_{\bm{r}_0}(\bm{r}) $ associated with the ion at $\bm{r}_0$), and hence quantum fluctuation, as well as the shift of the ion positions.
We assume that 
(i) in the bulk, all components of $p_i^{\bm{r}_0}$ and $q_{ij}^{\bm{r}_0}$ vanish due to the octahedral symmetry, 
(ii) near the $z=0$ surface, only $p_z^{\bm{r}_0}$ and $q_{zz}^{\bm{r}_0}$ are nonzero due to the four-fold rotation summery about the $z$ axis, 
(iii) near the $x=y=0$ hinge, $\bm{p}^{\bm{r}_0}$ points in the direction $(1,1,0)$ due to  the two-fold rotation symmetry about the (1,1,0) axis, and 
(iv) the nonzero components of dipoles and quadrupoles obey the lattice translation symmetry of the surface and the hinge. 
The dipole moments and the quadrupole moments of other surfaces and hinges are determined by the octahedral symmetry about the center of the cube. See Fig.~\ref{figqtilde}(b,c) for the illustration.

The electric field at the position $\bm{r}$ originating from these moments is given by
\begin{align}
&\delta\bm{E}(\bm{r})=\sum_{\bm{r}_0}\sum_{j}\frac{p_j^{\bm{r}_0}}{4\pi\epsilon_0}\frac{3r_ir_j-r^2\delta_{ij}}{r^5}\notag\\
&\quad\quad+\sum_{\bm{r}_0}\sum_{jk}\frac{q_{jk}^{\bm{r}_0}}{4\pi\epsilon_0}\frac{15r_ir_jr_k-3r^2(\delta_{ij}r_k+\delta_{ki}r_j)}{2r^7}.
\end{align}
By performing the summation under the assumptions summarized above, we find that the correction to $1/d^2$ term can be parametrized by a single parameter $\tilde{Q}$ as
\begin{align}
&\delta\bm{E}(0,0,d)=\chi_1\bm{n}_1\frac{\tilde{Q}}{4\pi\epsilon_0d^2}+O(d^{-3}),\label{surface001}\\
&\bm{n}_1\equiv\frac{(-1,-1,4)}{3\sqrt{2}},\quad\chi_1\equiv\frac{3\sqrt{2}}{4}=1.06066\cdots.
\end{align}
and
\begin{align}
&\delta\bm{E}(\tfrac{d}{\sqrt{3}},\tfrac{d}{\sqrt{3}},\tfrac{d}{\sqrt{3}})=\chi_2\bm{n}_2\frac{\tilde{Q}}{4\pi\epsilon_0d^2}+O(d^{-3}),\label{surface111}\\
&\bm{n}_2\equiv\frac{(1,1,1)}{\sqrt{3}},\quad \chi_2\equiv\frac{3(\sqrt{3}-1)}{2}=1.09808\cdots.
\end{align}
These expressions are valid in the intermediate length regime $a<d\ll L$; 
see the supplementary materials for the derivation of these results.
Note that the corrections are not isotropic and has $E_x=E_y$ components even for $\bm{r}=(0,0,d)$. This makes it possible to single out the pure corner charge contribution $Q_c$ in Eq.~\eqref{Fz1}.

To support these analytical results, we numerically investigate the contribution to the electric field around a corner from the surface relaxation illustrated in Fig.~\ref{figqtilde}(a).
We consider top three layers of a crystal and set $\Delta_1=0.07$~\AA, $\Delta_2=0.01$~\AA, $\Delta_3=0$, $d_1=2.76$~\AA, $d_2=2.80$~\AA, and $d_3=2.81$~{\AA} for the parameters in Fig.~\ref{figqtilde}(a)~\cite{VOGT2001155}. (For comparison, the bulk inter-layer distance is $2.80$ \AA~\cite{VOGT2001155}.)
The correction to the electric field due to the surface relaxation can be understood as the difference between the electric field from the relaxed structure and that from the unrelaxed one. This difference is equivalent to the electric field obtained by placing
 $\pm e$ point charges at the ionic positions after the relaxation, and those with the opposite charges at the lattice site before the relaxation.
Using Eq.~\eqref{eqB1} of Appendix~\ref{appbm}, we see that this configuration effectively corresponds to the value $\tilde{Q}=-0.036e$.
Our results are plotted in Fig.~\ref{figqtilde}(d,e), which are consistent with Eq.~\eqref{surface001}.
In the calculation, we took into account only the three surfaces and the three hinges adjacent to the focused corner.

We note that the value of $\tilde Q$ obtained numerically here assumes the point-charge model with charge $\pm e$ on each ion. From the analysis in Sec.\ \ref{sec:TB}, we expect quantum fluctuations to reduce $\tilde Q$ by about $10\%$ although it leaves the corner charge unchanged. This leads to a slightly better signal. In principle, there could also be higher-order corrections given the electron hopping integrals near the surface should also be modified due to boundary effects. An honest evaluation of such corrections require evaluating the {\it ab initio} charge density from a reasonably large finite sample that is open in all three spatial dimensions, which is beyond the scope of the present work.

\subsection{Direct observation by AFM measurement}
Having demonstrated the emergence of the fractional corner charge, here we propose an experiment for directly observing the predicted corner charge of $\mathrm{NaCl}$ by measuring the Coulomb force using an AFM.  
We suppose a cantilever tip with electric charge $Q_{\text{tip}}=+e$ is placed at the point $P$, at a distance $d>0$ above the corner $O$ as illustrated in Fig.~\ref{figcorner}(a).  The Coulomb force $\bm{F}=Q_{\text{tip}}\bm{E}(\bm{r})$ acting on the charged tip should be measured as a function of $d$.  Since the lattice constant is $a=5.60$ \AA~\cite{VOGT2001155}, the typical scale of the Coulomb force acting on the charged tip is
\begin{equation}
F_0\equiv \frac{e^2}{4\pi \epsilon_0 a^2}=7.36\times 10^{-10}\text{ }\mathrm{kg}\,\mathrm{m}\,\mathrm{s}^{-2}=736\text{ }\mathrm{pN}.
\end{equation}
For example, when $d=2a=11.2$ \AA, we get $F_z=F_0/32=23.0$ pN, which is a feasible number given AFM is capable of measuring pN \cite{Ohnesorge1451} or even sub-pN scale forces\cite{AOKI199745}. 
Note also that a larger electrostatic force can be achieved simply by increasing $Q_{\text{tip}}$. 
We remark that, the goal here is to measure the long-range electrostatic force from all the ions in the vicinity of the corner of the crystal, which gives rise to the emergent corner charge. This is very different from the problem of atomically resolving individual ions using AFM,~\cite{PhysRevLett.106.216102,PhysRevB.90.155455} which would typically rely on the short-range forces instead.~\cite{RevModPhys.75.949} In addition, instead of measuring the electrostatic force directly one could also measure the electric field distribution by combining scanning probe with quantum dots,~\cite{PhysRevLett.115.026101, PhysRevApplied.8.031002} which could enable the detection of the electric field over a much larger range. For instance, the mV-scale measurement in Ref.\ \onlinecite{PhysRevLett.115.026101} corresponds to a distance of $d/a \sim \mathcal O(100)$.

Importantly, the electric field originating from the corner charge is isotropic, whereas that from the distortion-induced dipole and higher surface moments is not.
In order to single out the contributions from the corner charge $Q_c$ from that coming from the surface distortion, one should measure the two components $E_x(\bm{r})$ and $E_z(\bm{r})$ separately as a function of $d$ (satisfying $a<d\ll L$) along the $z$ axis. The coefficient of the $1/d^2$ term of $E_x(\bm{r})$ determines the parameter $\tilde{Q}$ of the surface distortion in Eq.~\eqref{surface001}. One can subtract the effect of $\tilde{Q}$ from the $1/d^2$ term of $E_z(\bm{r})$ and obtain the pure corner charge contribution in Eq.~\eqref{Fz1}. One can further confirm the validity of this protocol by measuring the electric field along the $(1,1,1)$ direction, for which one expects that the contributions from $Q_c$ and $\tilde{Q}$ simply add up [Eq.~\eqref{surface111}].

\section{Discussion}
In this work, we showed that sodium chloride is an example of a three-dimensional octupole insulator with a fractional corner charge of $\pm e/8$. We further argue that the fractional charge could be directly observed using atomic force microscopy, and that the presence of defects in the crystal does not affect the asserted observability. Although surface relaxation induces non-zero surface dipole and higher moments which makes such a measurement nontrivial, we demonstrate that the intrinsic contribution from the fractionally quantized corner charge could be extracted thanks to its isotropic nature.
We further note that while our results imply the electrostatic signature of fractional corner charges should be measurable, the actual experiment is likely very challenging. In particular, we acknowledge that it could be nontrivial to accurately locate the corner of a three-dimensional sample using a scanning probe.
Yet, AFM is already routinely used to characterize two-dimensional materials to identify, for instance, the edges and corners of a monolayer terrace on a larger sample.~\cite{Pawlak_2012} As such, a step edge on the top of a flat surface (Fig.\ \ref{figexp}), which behaves like a two-dimensional crystal of sodium chloride and hosts fractional corner charges of $\pm e/4$, may be easier to observe experimentally.

The electrostatic-force measurement proposed in this work could also be used to measure the absolute value of the polarization of a one-dimensional insulator. Historically, the spontaneous polarization has been measured through a hystereses measurement~\cite{vanderbilt2018}: One applies a time-dependent electric field to the sample and measures the electric current $I(t)$ that flows through the sample. Assuming that the polarization completely flips during this process, the spontaneous polarization is determined by $\tilde{p}=(a/2)\int dt I(t)$. In contrast, our proposed AFM measurement determines the absolute value of the edge charge $Q$, whose fractional part is determined by the bulk polarization $p$ as illustrated in Fig.~\ref{polyA}. The difference of $p$ and $\tilde{p}$ may not be only an integer multiple of $ea$. (See the discussion of the ``formal" and the ``effective" polarization in Ref.~\cite[Sec.~4.4.2]{vanderbilt2018}).
For three-dimensional samples, even if the bulk dipole or quadrupole moments are nonzero, their associated boundary signatures may not be observable because of the large energy cost in sustaining a nonzero surface or hinge charge density; for such systems one expects significant surface reconstruction which neutralize the extended boundary charges. In contrast, for the ``top" moment in each dimension (i.e., the dipole moment in 1D, quadrupole moment in 2D, and the octupole moment in 3D) the associated boundary charges are localized to the corners, and this renders them directly measurable.

As we have demonstrated in Sec.\ \ref{sec:TB}, in a quantum mechanical treatment the effective charge on an ion is not generally an integer multiple of $e$. In molecular dynamics simulation such effects are commonly accounted for using the concept of ``partial charges'' \cite{ComputationalChem}, e.g., in modeling $\mathrm{Na}\mathrm{Cl}$  one may choose to assign  charges $\pm \delta$ with $\delta/e<1$ to the sodium and chlorine ions. 
From this perspective, our discussion amounts to a new class of higher-order constraints on the partial charge assignments, which demands that not only is  the total charge of the system quantized to an integer multiple of $e$, but also the lower-dimensional boundary charges associated to the surfaces (dipole), hinges (quadrupole) and corners (octupole). It is an interesting open question whether or not this realization could have any implications to the molecular dynamics in, say, salt formation and dissolution.

We also remark that, in this work, we have chosen to focus exclusively on $\mathrm{Na}\mathrm{Cl}$ for it exemplifies the key ideas and predictions underlying the connection of multipole insulators to ionic compounds. Our discussion applies immediately to other ionic compounds with the same formal oxidation numbers and structure, like $\mathrm{K}\mathrm{Cl}$, $\mathrm{Ag}\mathrm{F}$, to name but a few. In particular, from the perspective of effective charge models these different compounds come with very different bulk partial charges, but their corner charges are universally quantized to $\pm e/8$.  Furthermore, our results suggest that many common ionic compounds could be multipole insulators, and this opens up a new avenue for the experimental investigation of the implications and applications of nontrivial multipole moments in solids. It will also be interesting to explore if similar connections can be pursued for covalent solids.

\begin{acknowledgements}
We would like to thank Dominic Else, Berthold J{\"a}ck, Ethan Lake, Kam Tuen Law, Yukitoshi Motome, Shuichi Murakami, Naoto Nagaosa, Tai Kai Ng, Ding Pan, Shinichi Seki, Yoshiaki Sugimoto, and Senthil Todadri for useful discussions.
We thank Yusuke Nomura for kindly providing the tight-biding parameters derived from first-principles calculations.
The work of H.W. is supported by JSPS KAKENHI Grant No.~JP20H01825 and JP21H01789 and by JST PRESTO Grant No.~JPMJPR18LA. 
The work of H.C.P. is partly supported by a Pappalardo Fellowship at MIT and a Croucher Foundation Fellowship. 
\end{acknowledgements}

\appendix

\section{Review of fractional edge charge in one dimension
\label{app:1DNaCl}}
\begin{figure}
\begin{center}
\includegraphics[width=\columnwidth]{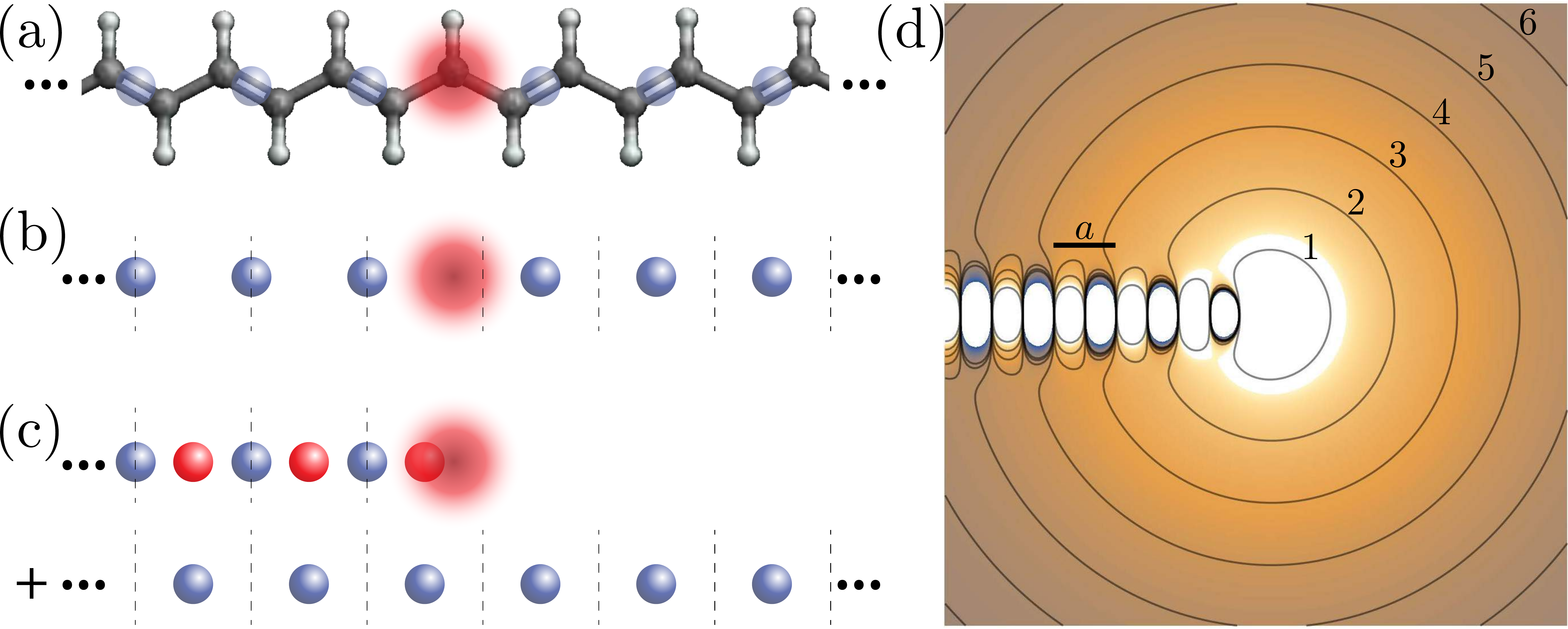}
\caption{\label{polyA} 
(a) Polyacetylene $(\mathrm{C}\mathrm{H})_n$ features an alternation between single and double covalent bonds along the carbon backbone. The domain wall traps an additional effective orbital, which we assume to be unoccupied here. Such an unoccupied domain wall has an effective positive charge of $e$ (red blob), the elementary charge. (b) In a purely electronic description, the main difference between the two sides of the domain wall is the location of the double bond. This can be described as localizing a pair of electrons (blue spheres) to either the boundary or the center of the unit cell (indicated by dashed line). (c) By introducing holes (red spheres; each of them representing a pair of holes as it is the ``inverse'' of a blue sphere), we can rewrite the charge distribution in (b) into the sum of a semi-infinite chain of alternating point charges and an infinite periodic chain. The  charge bound to the domain wall emerges at the end-point of the semi-infinite chain. (d) Electrostatic potential of a semi-infinite chain of alternating point charges of charge $\pm q$. In the polyacetylene example $q = 2e$. The equipotential lines at $(q/2)/(4 \pi \epsilon_0 \, n a)$ for $n=1,2,\dots, 6$ are shown and labeled by $n$, which confirms the emergence of an edge charge $q/2$. 
}
\end{center}
\end{figure}

In this appendix, we recapitulate how similar fractional charges arise in the conducting polymer polyacetylene  $(\mathrm{C}\mathrm{H})_n$~\cite{jackiw1976, su1979, heeger1988}.
The structure of polyacetylene is shown in Fig.~\ref{polyA}(a), which consists of alternating single and double covalent bonds between the atoms in the carbon backbone. Such alternation can be viewed as a spontaneous dimerization for which pairs of carbon atoms become more strongly bonded to each other. Importantly, there are two patterns of dimerization differing only by how the atoms are grouped. In an infinite chain with a uniform dimerization pattern, the system is an electrical insulator. However, if there is a domain wall in the system separating regions of distinct dimerization patterns, there is an additional effective orbital localized to the domain wall. With exact chiral symmetry the energy of the orbital is pinned to zero, but realistic perturbations can move the orbital away from the chemical potential. For the purpose of our discussion, we only consider spin-rotation invariant states, and as such the orbital could only be empty or doubly filled. If the orbital is empty, the atoms in the vicinity of the domain wall collectively lose one electron, which implies the empty domain wall has charge $e$, where $e$ denotes the elementary charge. If the orbital is doubly filled, effectively an additional electron is bound to the domain wall, and thus it has charge $-e$. 
 These charged domain walls are in fact mobile, which render the polymer conducting.

From a purely electronic perspective, one can focus only on the excess electrons which contribute to the double bonds in the carbon backbone. As shown in Fig.~\ref{polyA}(b), these electrons can be modeled as being localized to the center of the double bonds. Once a unit cell convention is picked, the two dimerization patterns can be differentiated by whether the electrons are localized to the boundary ($\tilde x=0$) or the center ($\tilde x=1/2$) of the 1D unit cell, where $\tilde x \in [0,1)$ denotes dimensionless coordinates within the unit cell. In addition, for an infinite chain with a uniform dimerization pattern the system has an inversion symmetry about either of $\tilde x=0$ or $\tilde x=1/2$. Although the choice of the unit cell for a uniform chain is {\it ad hoc}, the relative shift of the electron position across the two sides of the domain wall is well-defined, and the trapping of an additional effective electronic orbital at the domain wall is a consequence of such a shift. Furthermore, notice that the system has spin rotation symmetry, and so the electron count is effectively doubled (i.e., there are two more electrons in a double bond compared to a single bond). The domain wall, however, carries charge $\pm e$ depending on whether it is empty or doubly occupied, and as such the charge it traps is half of the basic charge unit in the problem. In this sense, we say the charge of the domain wall is fractional.

Next we relate the fractional charges in polyacetylene to fractional edge charges in a 1D ionic crystal. This could be achieved by filling the $\tilde x=1/2$ positions to the left of the domain wall simultaneously by a pair of electrons and a pair of holes. 
With this formal rewriting, we can now regroup the added electron and hole pairs in the way shown in Fig.~\ref{polyA}(c):
The added electron pairs can be grouped with those originally localized to the right of the domain wall, which combines into an infinite 1D chain with a uniform dimerization pattern; 
the holes we introduced can be grouped with the electrons originally localized to the $\tilde x=0$ positions to the left of the domain wall, which gives a semi-infinite chain of alternating point charges which end at the domain wall. 
Given the chain with a uniform dimerization pattern has lattice translation invariance, the fractional charge bound to the domain wall has to be assigned to the end of the semi-infinite chain of alternating point charges. This establishes the connection we asserted. 

The physical picture above can also be verified explicitly by considering a model of alternating point charges of charge $\pm q$. 
We consider a semi-infinite system of point charges specified by the linear charge density
\begin{equation}\begin{split}\label{eq:1DNaCl}
\lambda (x) = q \sum_{n=0}^{\infty}  (-1)^n \delta(x + n a/2),
\end{split}\end{equation}
where $a$ is the lattice constant.
Note that this recovers the one-dimensional version of NaCl by setting $q=e$.  
To clarify, we consider a one-dimensional chain embedded in the three-dimensional space, such that the electric potential at a distance $r$ away from a point charge of charge $q$ is $q/4 \pi \epsilon_0 r$. In this regard the delta functions in the other two coordinates are implicit in the charge density above.
In the following we solve for the electrostatic potential $V$ generated by the semi-infinite chain of alternating point charges defined in Eq.~\eqref{eq:1DNaCl}.

Let us adopt cylindrical coordinates with the longitudinal axis along $x$ such that a point in space is labeled by $(\rho, \varphi, x)$. The system has azimuthal symmetry and so there will be no dependence on the angle $\varphi$.
Along the positive $x$-axis, i.e., for $\rho=0$ and $x>0$, the infinite series can be related to the digamma function $\psi$ ~\cite[Eq.~5.7.7]{DLMF}:
\begin{equation}\begin{split}\label{eq:}
V(0,x) =& \frac{q}{4 \pi \epsilon_0} \sum_{n=0}^{\infty}  (-1)^n \frac{1}{x+ a n/2}\\
 =& \frac{q}{4 \pi \epsilon_0 a}
  \left(   \psi \left(\frac{x}{a}+\frac{1}{2} \right) - \psi\left(\frac{x}{a}  \right)\right),
\end{split}\end{equation}
where $\psi(x)$ is related to the gamma function $\Gamma(x)$ by $\psi(x) = \Gamma'(x)/\Gamma(x) = \frac{d}{dx} \ln \Gamma(x)$. For (real) $\tilde x>0$ the digamma function $\psi(\tilde x)$ has an integral representation~\cite[Eq.~5.9.12]{DLMF}
\begin{equation}\begin{split}\label{eq:}
\psi(\tilde x) = \int_0^\infty d \tilde k \left( \frac{e^{-\tilde k}}{\tilde k} - \frac{e^{- \tilde k \tilde x}}{1-e^{-\tilde k}}\right),
\end{split}\end{equation}
 where the ornamentation $\sim$ indicates dimensionless variables. This gives
\begin{equation}\begin{split}\label{eq:}
\psi(\tilde x+1/2)-\psi(\tilde  x) =   \int_0^\infty d\tilde k \frac{e^{- \tilde k \tilde x}}{e^{-\tilde k/2}+1}.
\end{split}\end{equation}
Comparing with the general solution of the Laplace's equation in our setting, we can infer that the electrostatic potential at general values of $(\rho, x> 0)$ is
\begin{equation}\begin{split}\label{eq:CylindricalV}
V(\rho, x) = \frac{q}{4\pi \epsilon_0}  \int_0^\infty dk \, J_0(k \rho) \frac{e^{- k x}}{e^{-k a/2}+1}.
\end{split}\end{equation}

To see the emergence of a half charge at the edge, we consider a multipole expansion of $V(0,x>0)$ and examine the coefficient of the leading monopole term. 
The expansion could be obtained by first redefining $\tilde k = k x$ and then consider the formal series of the integrand in powers of $a/x$:
\begin{equation}\begin{split}\label{eq:Vasym}
V(0, x>0)  =&\frac{q}{4\pi \epsilon_0 x }  \int_0^\infty d \tilde k \, \frac{e^{- \tilde k }}{e^{- \tilde k a/(2 x)}+1}\\
 =&\frac{q}{4\pi \epsilon_0 x }  \int_0^\infty d \tilde k \, e^{- \tilde k }
\sum_{n=0}^\infty \frac{E_{n}(0)}{2 \, n!} \left( -\frac{\tilde k a}{2 x}\right)^n\\
 =&\frac{q}{4\pi \epsilon_0 x } 
\sum_{n=0}^\infty \frac{E_{n}(0)}{2} \left( -\frac{ a}{2 x}\right)^n\\
 =&\frac{q}{4\pi \epsilon_0 x } \left( \frac{1}{2} -
\sum_{m=0}^\infty \frac{E_{2m+1}(0)}{2} \left( \frac{ a}{2 x}\right)^{2m+1} \right),
\end{split}\end{equation}
where $E_{n}(0)$ denotes the Euler polynomial evaluated at $0$,~\cite[Sec.~24.2]{DLMF} and in the last line we used $E_{2m}=0$ for all $m\geq 1$. More explicitly, the lowest few non-vanishing terms of the expansion read
\begin{equation}\begin{split}\label{eq:}
V(0, x>0) 
 =&\frac{q}{4\pi \epsilon_0 x } \left( \frac{1}{2} +  \frac{ a}{8 x} -   \frac{ a^3}{64 x^3}+ \frac{a^5}{128 x^5} + \dots\right),
\end{split}\end{equation}
and we see explicitly that the leading (monopole) term corresponds to a charge $q/2$ placed at the origin.

The asymptotic expansion above also allows us to directly write down the general form of $V$ in spherical coordinates. Let $\rho = r \sin \theta $ and $x = r \cos \theta $ with $\theta < \pi/2$. To obtain $V(r,\theta)$, we simply replace $x\mapsto r$ and attach the Legendre polynomial $P_{2m+1}(\cos\theta)$ to the $(a/x)^{2m+1}$ term in Eq.~\eqref{eq:Vasym}. 
Although for finite value of $a/x$ the expansion does not converge due to the rapidly growing factor $E_{2m+1}(0)$, it allows one to bypass the oscillatory behavior of the integrand in Eq.~\eqref{eq:CylindricalV}, and could be used to obtain a well-behaved estimate of the potential at any point in the half-space $x>0$ through an optimal truncation of the asymptotic series. As a technical trick, more accurate numerical estimates could be obtained by keeping the contribution from the point charges in the interval $x \in (L,0]$ for some finite $L \gg a$, and then sum the contribution from all the remaining charges into the asymptotic series discussed above. 
The plot in Fig.~\ref{polyA}(d) is obtained with $L = 10 a$ together with an optimal truncation of the asymptotic series, which confirms the electric field near the edge of the semi-infinite chain indeed correspond to that of a fractional edge charge $q/2$.

We note in passing that the analysis here is closely related to that for a point charge within a parallel-plate capacitor, which begets as an image problem an infinite chain of alternating point charges~\cite{HEUBRANDTNER2002439}.

\section{Electric field contribution from boundary moments}
\label{appbm}
In this appendix, we discuss the contribution from hinges and surfaces to the $1/d^2$ term of the electric field around a corner. We assume the following form of the dipole moments and quadrupole moments [see Fig.~\ref{figqtilde}(b,c)]:
The $x=y=0$ hinge has the dipole moment $\bm{p}=(p_h,p_h,0)$ per unit length along the hinge.
The $z=0$ surface has a dipole moment $\bm{p}=(0,0,p_s^{\bm{r}_0})$ at the position $\bm{r}=-\bm{r}_0-(n,m,0)$ in each surface unit cell spanned by $(-1,0,0)$ and $(0,-1,0)$.
Here, $\bm{r}_0=(x_0,y_0,z_0)$ represents the relative position within the surface unit cell. We consider three cases (a) $\bm{r}_0=(0,0,z_0)$, (b) $\bm{r}_0=(1/2,1/2,z_0)$, and (c) the pair of $\bm{r}_0=(1/2,0,z_0)$ and $(0,1/2,z_0)$, assuming the four-fold rotation symmetry.
The dipole moments on these points are $p_s^{a,z_0}$, $p_s^{b,z_0}$, and $p_s^{c,z_0}$, respectively.
The surface also has the quadrupole moment $q_{ij}=\delta_{iz}\delta_{jz}q_s$ per unit area. The moments on other hinges and surfaces are determined by the octahedral symmetry as illustrated in Fig.~\ref{figqtilde}(b,c).
In this setting, $\tilde{Q}$ in Eqs.~\ref{surface001} and \eqref{surface111} in the limit of large $L$ is given by
\begin{align}
\tilde{Q}=&\sum_{z_0}\Big[(1+2z_0)\frac{p_s^{a,z_0}}{a}+(2z_0)\frac{p_s^{b,z_0}}{a}+(1+4z_0)\frac{p_s^{c,z_0}}{a}\Big]\notag\\
&\quad\quad+\frac{2p_h}{a}-\frac{q_s}{a^2}.\label{eqB1}
\end{align}
In the following, we present the detailed derivation of this result.
For brevity, we set $a=1$ and neglect $O(d^{-3})$ terms of the electric field.

\subsection{Hinge dipole moments}
Let us start with hinge dipole moments. We focus on the hinge at $x=y=0$ first. We assume that the hinge is semi-infinite and fills the $z\leq0$ region of the $z$ axis [see Fig.~\ref{figqtilde}(c)].
The $1/d^2$ term of the electric field produced by the dipole moment density $\bm{p}=(p_h,p_h,0)$ is given by the sum
\begin{align}
&\bm{E}_{\text{$z$-axis}}^{\bm{p}}(\bm{r})=\sum_{n=0}^{\infty}\bm{E}^{\bm{p}}(\bm{r}+(0,0,n)),\\
&E_i^{\bm{p}}(\bm{r})\equiv\sum_{j}\frac{p_j}{4\pi\epsilon_0}\frac{3r_ir_j-r^2\delta_{ij}}{r^5}.
\end{align}
We approximate the sum by an integral as
\begin{align}
\sum_{n=0}^{\infty}\bm{E}^{\bm{p}}(\bm{r}+(0,0,n))\simeq\int_{0}^{\infty}dn\bm{E}^{\bm{p}}(\bm{r}+(0,0,n)).
\end{align}
For example, for the observing points $\bm{r}=(d,0,0)$, $(0,d,0)$, $(0,0,d)$, and $(d/\sqrt{3},d/\sqrt{3},d/\sqrt{3})$, we find
\begin{align}
\bm{E}_{\text{$z$-axis}}^{(p_h,p_h,0)}(d,0,0)&=(1,-1,1)\frac{p_h}{4\pi \epsilon_0d^2},\\
\bm{E}_{\text{$z$-axis}}^{(p_h,p_h,0)}(0,d,0)&=(-1,1,1)\frac{p_h}{4\pi \epsilon_0d^2},\\
\bm{E}_{\text{$z$-axis}}^{(p_h,p_h,0)}(0,0,d)&=-\frac{1}{2}(1,1,0)\frac{p_h}{4\pi \epsilon_0d^2},\\
\bm{E}_{\text{$z$-axis}}^{(p_h,p_h,0)}(\tfrac{d}{\sqrt{3}},\tfrac{d}{\sqrt{3}},\tfrac{d}{\sqrt{3}})&=\frac{(9-5\sqrt{3},9-5\sqrt{3},4\sqrt{3})}{6}\frac{p_h}{4\pi \epsilon_0d^2}.
\end{align}

The other two hinges connected to the corner $O$ of the cube [see Fig.~\ref{figqtilde}(c)] support symmetry-related dipoles $\bm{p}=(0,p_h,p_h)$ along the $x$ axis and $\bm{p}=(p_h,0,p_h)$ along the $y$ axis.  The sum of the contributions from these three hinges is 
\begin{align}
\bm{E}_{\text{hinges}}^{p_h}(\bm{r})&\equiv\bm{E}_{\text{$x$-axis}}^{(0,p_h,p_h)}(\bm{r})+\bm{E}_{\text{$y$-axis}}^{(p_h,0,p_h)}(\bm{r})+\bm{E}_{\text{$z$-axis}}^{(p_h,p_h,0)}(\bm{r}).
\end{align}
We find
\begin{align}
\bm{E}_{\text{hinges}}^{p_h}(0,0,d)&=\chi_1\bm{n}_1\frac{2p_h}{4\pi \epsilon_0d^2}
\end{align}
and
\begin{align}
\bm{E}_{\text{hinges}}^{p_h}(\tfrac{d}{\sqrt{3}},\tfrac{d}{\sqrt{3}},\tfrac{d}{\sqrt{3}})&=\chi_2\bm{n}_2\frac{2p_h}{d^2}.
\end{align}
Therefore, the hinge dipole moment $p_h$ gives the contribution $2p_h/a$ to $\tilde{Q}$.

\subsection{Surface quadrupole moments}
Next let us discuss surface quadrupole moments. We focus on the $z=0$ surface first. We assume that the surface is semi-infinite and fills the $x\leq0$ and $y\leq0$ region of the $xy$ plane [see Fig.~\ref{figqtilde}(b)].

The $1/d^2$ term of the electric field produced by the dipole moment density $q_{zz}=q_s$ is given by the sum
\begin{align}
&\bm{E}_{\text{$xy$-plane}}^{q}(\bm{r})=\sum_{n,m=0}^{\infty}\bm{E}^{q}(\bm{r}+(n,m,0)),\\
&E_i^{q}(\bm{r})\equiv\sum_{jk}\frac{q_{jk}}{4\pi\epsilon_0}\frac{15r_ir_jr_k-3r^2(\delta_{ij}r_k+\delta_{ki}r_j)}{2r^7}.
\end{align}
We approximate the sum by an integral as
\begin{align}
&\sum_{n,m=0}^{\infty}\bm{E}^{q}(\bm{r}+(n,m,0))\notag\\
&\simeq\int_{0}^{\infty}dn\int_{0}^{\infty}dm\bm{E}^{q}(\bm{r}+(n,m,0)).
\end{align}
For example, for the observing points $\bm{r}=(d,0,0)$, $(0,d,0)$, $(0,0,d)$, and $(d/\sqrt{3},d/\sqrt{3},d/\sqrt{3})$, we find
\begin{align}
\bm{E}_{\text{$xy$-plane}}^{q_{zz}=q_s}(d,0,0)&=-\frac{1}{4}(2,1,0)\frac{q_s}{4\pi \epsilon_0d^2},\\
\bm{E}_{\text{$xy$-plane}}^{q_{zz}=q_s}(0,d,0)&=-\frac{1}{4}(1,2,0)\frac{q_s}{4\pi \epsilon_0d^2},\\
\bm{E}_{\text{$xy$-plane}}^{q_{zz}=q_s}(0,0,d)&=\frac{1}{2}(1,1,0)\frac{q_s}{4\pi \epsilon_0d^2},\\
\bm{E}_{\text{$xy$-plane}}^{q_{zz}=q_s}(\tfrac{d}{\sqrt{3}},\tfrac{d}{\sqrt{3}},\tfrac{d}{\sqrt{3}})&=-\frac{(\sqrt{3},\sqrt{3},18-8\sqrt{3})}{12}\frac{q_s}{4\pi \epsilon_0d^2}.
\end{align}
Summing up contributions from other two symmetry-related surfaces, we obtain
\begin{align}
\bm{E}_{\text{surfaces}}^{q_s}(\bm{r})\equiv\bm{E}_{\text{$yz$-plane}}^{q_{xx}=q_s}(\bm{r})+\bm{E}_{\text{$zx$-plane}}^{q_{yy}=q_s}(\bm{r})+\bm{E}_{\text{$xy$-plane}}^{q_{zz}=q_s}(\bm{r}).
\end{align}
We find
\begin{align}
\bm{E}_{\text{surfaces}}^{q_s}(0,0,d)=-\chi_1\bm{n_1}\frac{q_s}{4\pi \epsilon_0d^2}
\end{align}
and
\begin{align}
\bm{E}_{\text{surfaces}}^{q_s}(\tfrac{d}{\sqrt{3}},\tfrac{d}{\sqrt{3}},\tfrac{d}{\sqrt{3}})=-\chi_2\bm{n}_2\frac{q_s}{4\pi \epsilon_0d^2}.
\end{align}
Therefore, the surface quadrupole moment $q_s$ gives the contribution $-q_s/a^2$ to $\tilde{Q}$.

\subsection{Surface dipole moments}
Finally let us discuss surface dipole moments. 
An extra care is needed for the discussion of the surface dipole moments, as naively the leading order term is $1/d$, not $1/d^2$. We will see that the $1/d$ term is cancelled after summing up contributions from three symmetry-related surfaces.

We focus on the $z=0$ surface. The $1/d^2$ term of the electric field produced by the dipole moments on this surface is given by the sum
\begin{align}
\bm{E}_{\text{$xy$-plane}}^{\bm{p},\bm{r}_0}(\bm{r})=\sum_{n,m=0}^{\infty}\bm{E}^{\bm{p}}(\bm{r}+\bm{r}_0+(n,m,0)).
\end{align}
To perform the summation correctly taking into account the $O(d^{-2})$ terms of the electric field, we use the Euler--Maclaurin formula:
\begin{align}
&\sum_{n,m=0}^{\infty}\bm{E}^{\bm{p}}(\bm{r}+\bm{r}_0+(n,m,0))\notag\\
&=\int_{0}^\infty dn \int_{0}^{\infty}dm\bm{E}^{\bm{p}}(\bm{r}+\bm{r}_0+(n,m,0))\notag\\
&\quad+\frac{1}{2}\int_{0}^{\infty}dn\bm{E}^{\bm{p}}(\bm{r}+\bm{r}_0+(n,0,0))\notag\\
&\quad+\frac{1}{2}\int_{0}^{\infty}dm\bm{E}^{\bm{p}}(\bm{r}+\bm{r}_0+(0,m,0)).
\end{align}
For $\bm{r}_0=(0,0,z_0)$, we find
\begin{align}
&\bm{E}_{\text{$xy$-plane}}^{(0,0,p_s^{a,z_0})}(d,0,0)\notag\\
&=-(0,0,1)\frac{p_s^{a,z_0}}{4\pi \epsilon_0d}+\frac{1}{4}(4z_0,2z_0,-3)\frac{p_s^{a,z_0}}{4\pi \epsilon_0d^2},
\end{align}
\begin{align}
&\bm{E}_{\text{$xy$-plane}}^{(0,0,p_s^{a,z_0})}(0,d,0)\notag\\
&=-(0,0,1)\frac{p_s^{a,z_0}}{4\pi \epsilon_0d}+\frac{1}{4}(2z_0,4z_0,-3)\frac{p_s^{a,z_0}}{4\pi \epsilon_0d^2},
\end{align}
\begin{align}
&\bm{E}_{\text{$xy$-plane}}^{(0,0,p_s^{a,z_0})}(0,0,d)\notag\\
&=(1,1,0)\frac{p_s^{a,z_0}}{4\pi \epsilon_0d}+\frac{1}{4}(2-4z_0,2-4z_0,4)\frac{p_s^{a,z_0}}{4\pi \epsilon_0d^2},
\end{align}
\begin{align}
&\bm{E}_{\text{$xy$-plane}}^{(0,0,p_s^{a,z_0})}(\tfrac{d}{\sqrt{3}},\tfrac{d}{\sqrt{3}},\tfrac{d}{\sqrt{3}})\notag\\
&=\frac{\sqrt{3}-1}{2}(1,1,-2)\frac{p_s^{a,z_0}}{4\pi\epsilon_0d}\notag\\
&\quad+\frac{(9-2\sqrt{3},9-2\sqrt{3},-2\sqrt{3})}{12}\frac{p_s^{a,z_0}}{4\pi\epsilon_0d^2}\notag\\
&\quad+\frac{2z_0(\sqrt{3},\sqrt{3},18-8\sqrt{3})}{12}\frac{p_s^{a,z_0}}{4\pi\epsilon_0d^2}.
\end{align}
For $\bm{r}_0=(1/2,1/2,z_0)$, we find
\begin{align}
&\bm{E}_{\text{$xy$-plane}}^{(0,0,p_s^{b,z_0})}(d,0,0)\notag\\
&=-(0,0,1)\frac{p_s^{b,z_0}}{4\pi \epsilon_0d}+\frac{1}{4}(4z_0,2z_0,0)\frac{p_s^{b,z_0}}{4\pi \epsilon_0d^2},
\end{align}
\begin{align}
&\bm{E}_{\text{$xy$-plane}}^{(0,0,p_s^{b,z_0})}(0,d,0)\notag\\
&=-(0,0,1)\frac{p_s^{b,z_0}}{4\pi \epsilon_0d}+\frac{1}{4}(2z_0,4z_0,0)\frac{p_s^{b,z_0}}{4\pi \epsilon_0d^2},
\end{align}
\begin{align}
&\bm{E}_{\text{$xy$-plane}}^{(0,0,p_s^{b,z_0})}(0,0,d)\notag\\
&=(1,1,0)\frac{p_s^{b,z_0}}{4\pi \epsilon_0d}+\frac{1}{4}(-4z_0,-4z_0,0)\frac{p_s^{b,z_0}}{4\pi \epsilon_0d^2},
\end{align}
\begin{align}
&\bm{E}_{\text{$xy$-plane}}^{(0,0,p_s^{b,z_0})}(\tfrac{d}{\sqrt{3}},\tfrac{d}{\sqrt{3}},\tfrac{d}{\sqrt{3}})\notag\\
&=\frac{\sqrt{3}-1}{2}(1,1,-2)\frac{p_s^{b,z_0}}{4\pi\epsilon_0d}\notag\\
&\quad+\frac{2z_0(\sqrt{3},\sqrt{3},18-8\sqrt{3})}{12}\frac{p_s^{b,z_0}}{4\pi\epsilon_0d^2}.
\end{align}
Finally, for $\bm{r}_0=(1/2,0,z_0)$ and $(0,1/2,z_0)$, we find
\begin{align}
&\bm{E}_{\text{$xy$-plane}}^{(0,0,p_s^{c,z_0})}(d,0,0)\notag\\
&=-(0,0,2)\frac{p_s^{c,z_0}}{4\pi \epsilon_0d}+\frac{1}{4}(8z_0,4z_0,-3)\frac{p_s^{c,z_0}}{4\pi \epsilon_0d^2},
\end{align}
\begin{align}
&\bm{E}_{\text{$xy$-plane}}^{(0,0,p_s^{c,z_0})}(0,d,0)\notag\\
&=-(0,0,2)\frac{p_s^{c,z_0}}{4\pi \epsilon_0d}+\frac{1}{4}(4z_0,8z_0,-3)\frac{p_s^{c,z_0}}{4\pi \epsilon_0d^2},
\end{align}
\begin{align}
&\bm{E}_{\text{$xy$-plane}}^{(0,0,p_s^{c,z_0})}(0,0,d)\notag\\
&=(2,2,0)\frac{p_s^{c,z_0}}{4\pi \epsilon_0d}+\frac{1}{2}(2-8z_0,2-8z_0,4)\frac{p_s^{c,z_0}}{4\pi \epsilon_0d^2},
\end{align}
\begin{align}
&\bm{E}_{\text{$xy$-plane}}^{(0,0,p_s^{c,z_0})}(\tfrac{d}{\sqrt{3}},\tfrac{d}{\sqrt{3}},\tfrac{d}{\sqrt{3}})\notag\\
&=\frac{\sqrt{3}-1}{2}(2,2,-4)\frac{p_s^{c,z_0}}{4\pi\epsilon_0d}\notag\\
&\quad+\frac{(9-2\sqrt{3},9-2\sqrt{3},-2\sqrt{3})}{12}\frac{p_s^{c,z_0}}{4\pi\epsilon_0d^2}\notag\\
&\quad+\frac{4z_0(\sqrt{3},\sqrt{3},18-8\sqrt{3})}{12}\frac{p_s^{c,z_0}}{4\pi\epsilon_0d^2}.
\end{align}
We combine contributions from other two symmetry-related surfaces:
\begin{align}
\bm{E}_{\text{surfaces}}^{p_s}(\bm{r})&\equiv\sum_{w=a,b,c}\sum_{z_0}\Big[\bm{E}_{\text{$yz$-plane}}^{(p_s^{w,z_0},0,0)}(\bm{r})\notag\\
&\quad+\bm{E}_{\text{$zx$-plane}}^{(0,p_s^{w,z_0},0)}(\bm{r})+\bm{E}_{\text{$xy$-plane}}^{(0,0,p_s^{w,z_0})}(\bm{r})\Big].
\end{align}
We find
\begin{align}
&\bm{E}_{\text{surfaces}}^{p_s^w}(0,0,d)\notag\\
&=\chi_1\bm{n}_1\frac{\sum_{z_0}[(1+2z_0)p_s^{a,z_0}+(2z_0)p_s^{b,z_0}+(1+4z_0)p_s^{c,z_0}]}{4\pi \epsilon_0d^2}
\end{align}
and
\begin{align}
&\bm{E}_\text{surfaces}^{p_s}(\tfrac{d}{\sqrt{3}},\tfrac{d}{\sqrt{3}},\tfrac{d}{\sqrt{3}})\notag\\
&=\chi_2\bm{n}_2\frac{\sum_{z_0}[(1+2z_0)p_s^{a,z_0}+(2z_0)p_s^{b,z_0}+(1+4z_0)p_s^{c,z_0}]}{4\pi\epsilon_0d^2}.
\end{align}
Therefore, the surface dipole moment $p_s$ gives the contribution $\sum_{z_0}[(1+2z_0)p_s^{a,z_0}+(2z_0)p_s^{b,z_0}+(1+4z_0)p_s^{c,z_0}]/a$. 

\bibliography{ref}

\clearpage

\end{document}